\documentclass[sn-basic, twocolumn]{sn-jnl}

\usepackage{hyperref}
\usepackage{fixltx2e}
\usepackage{graphicx}
\usepackage{longtable}
\usepackage{float}
\usepackage{wrapfig}
\usepackage{rotating}
\usepackage{dcolumn}
\usepackage{lineno}
\usepackage[normalem]{ulem}
\usepackage{amsmath}
\usepackage{mathtools}
\usepackage{textcomp}
\usepackage{marvosym}
\usepackage{wasysym}
\usepackage{multicol}
\usepackage{amssymb}
\usepackage{verbatim}
\usepackage{threeparttable}
\usepackage{booktabs}
\usepackage{textcomp}
\usepackage{cuted}
\author*[1]{\fnm{Weiliang} \sur{Tao}}\email{taoweiliang@whu.edu.cn}
\affil*[1]{\orgdiv{Electronic Information School}, \orgname{Wuhan University}, \orgaddress{\state{Wuhan}, \country{China}}}
\author[2]{\fnm{Yan} \sur{Liu}}\email{lyly9633@sina.com}
\affil[2]{\orgdiv{State Key Laboratory of Power Grid Environmental Protection}, \orgname{China Electric Power Research Institute}, \orgaddress{\state{Wuhan}, \country{China}}}
\author[1]{\fnm{Zhimin} \sur{Ma}}\email{whumazm@163}
\author[1]{\fnm{Wenbin} \sur{Hu}}\email{WenbinHu@whu.edu.cn}
\date{}
\title{Two-dimensional flow field measurement of sediment-laden flow based on ultrasound image velocimetry}
\begin{document}

\abstract{This paper proposes a novel particle image velocimetry (PIV) technique to generate an instantaneous two-dimensional velocity field for sediment-laden fluid based on optical flow algorithm of ultrasound imaging.  In this paper, an ultrasonic PIV (UPIV) system is constructed by integrating a medical ultrasound instrument and an ultrasonic particle image velocimetry algorithm.  The medical ultrasound instrument with a phased sensor array is used to acquire acoustic echo signals and generate two-dimensional underwater ultrasound images.  Based on the optical flow field, the instantaneous velocity of the particles in water corresponding to the pixels in the ultrasonic particle images is derived from the grayscale change between adjacent images under the L-K local constraint, and finally the two-dimensional flow field is obtained.  Through multiple sets of experiments, the proposed algorithm is verified.  The experimental results are compared with those of the conventional cross-correlation algorithms.  The results show that the L-K optical flow method can not only obtain the underwater velocity field accurately, but also has the advantages of good smooothness and extensive suitability, specially for the flow field measurement in sediment-laden fluid than conventional algorithms.}


\maketitle

\section{Introduction}
\label{sec:org7db1263}
The measurement of the depth-averaged velocity in sediment-laden flow is a prerequisite for many research works on river engineering, such as river-bank erosion, sediment transportation.  Two-dimension quantitative information of the depth-averaged velocity field helps investigate dynamic flow structures in complicated turbulent phenomena.  The information enables us to compute vorticity, deformation, and other quantities, which are directly related to the dynamics of coherent flow structures \cite{tang08_gener_model_later_depth_averag}.  However, traditional measuring instruments, such as hot-wires and laser-Doppler anemometers, are of one-point measurements, not capable of illustrating an instantaneous spatial flow structure.  The conventional Doppler algorithm has a problem on the angulation error, which confines the measurement only to the velocity component along ultrasonic beams.  Parallel alignment of the ultrasonic beam to the flow direction is required, which is not always feasible in many applications.  It is challenging research to scientists and hydraulic engineers for a long time \cite{byrd00_estim_depth_averag_veloc_rough}.

Optical PIV now is widely adopted as a reliable method that can obtain quantitative information on a spatial flow structure \cite{adrian91_partic_imagin_techn_exper_fluid_mechan}.  However, the light cannot penetrate through turbid water because it rapidly attenuates when traveling in turbid flows, which affects their applications in sediment-laden flow.

To overcome those disadvantages and meet requirements of high-precision measurements \cite{huang93_limit_improv_piv}, an alternative implementation based on ultrasound imaging is developed \cite{crapper00_flow_field_visual_sedim_laden}. A UPIV system is built by integrating a medical ultrasound instrument and a measurement algorithm of particle velocimetry with ultrasound images. The combination of ultrasound imaging technique and a digital PIV method can generate a new measurement technique of two-dimensional, two-component, depth-averaged velocity field appropriate for sediment-laden flow conditions \cite{poelma16_ultras_imagin_veloc}.  Compared with Doppler-based techniques, an advantage of the UPIV is its ability to measure the flow velocity components in both parallel and perpendicular directions to the ultrasonic beam.  In general, it has excellent performance on simplicity, accuracy, and accessibility, while it does not have the limitations of the Doppler-based algorithm \cite{demarchi12_echo_partic_image_veloc}.

\begin{figure}[htbp]
\centering
\includegraphics[width=.9\linewidth]{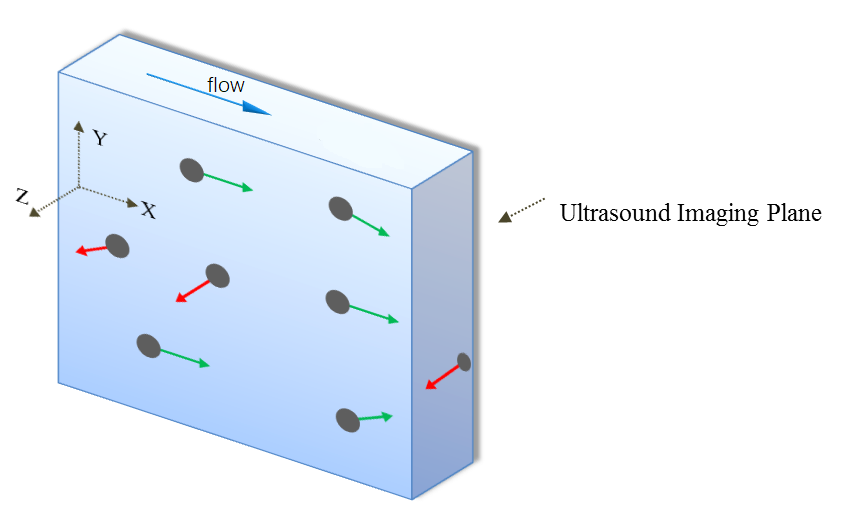}
\caption{\label{fig:orgb3b27a6}
the In-Plane Motion and Out-of-Plane Motion of particles}
\end{figure}

Even with ultrasonic particle image velocimetry, flow velocity measurement algorithms still faces a major challenge.  When the concentration of sandy water is high enough, algorithms based on the PTV method of tracking individual particles can no longer meet the measurement requirements, because there is not only In-Plane Motion but also Out-of-Plane Motion in the acoustic plane of ultrasound images, which brings instability to the calculation, as shown in Fig. \ref{fig:orgb3b27a6}.  On the other hand, there are a large number of particles in turbid water.  Their motion trajectories cross each other, whick causes uncertainty to the analysis of particles motion and greatly reduce the tracking robustness.

\begin{figure}[htbp]
\centering
\includegraphics[width=0.8\columnwidth]{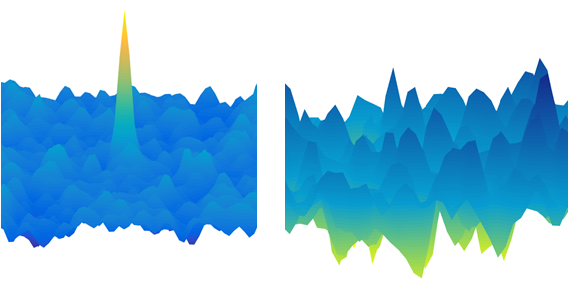}
\caption{\label{fig:org7b447a4}
Single peak and multiple peaks in cross-correlation plane}
\end{figure}

Although PIV matching algorithms based on cross-correlation method is continuously improved, such as the FFT algorithm is used to improve the real-time performance of the flow field calculation, the algorithms have some shortcommings as follows:

\begin{itemize}
\item According to the motion characteristics of high concentration sand-bearing flow, the main influences on the motion of sediment particles are the resistance caused by relative motions between particles, the upward force cased by shearing actions of the water flow, and the additional mass force caused by variable speed motion of the correlation of the B-type ultrasound image sequences acquired in the correlation calculation in the velocimetry of high concentration sandy water flow.

\item Because of the high concentration of tracer particles,  it will make the correlation map appear with multiple peaks.  As shown in Fig.\ref{fig:org7b447a4}
\end{itemize}

\begin{itemize}
\item From the normalized cross-correlation formula, it is known that the cross-correlation algorithms are based on stastical calculation, and the velocity vectors obtained by those algorithms are only the displacement vectors with the highest confidence in probability.
\end{itemize}

In recent years, particle image velocimetry algorithms based on the optical flow method have received increasing attention from researchers because they overcome some shortcommings described above \cite{kaehler16_main_resul_inter_piv_chall}.  The optical flow method, which considers the motion vector field as a whole, is suitable for the calculation and analysis of image sequence with highly concentrated particles and continuous deformation.  With dense optical flow algorithm, its resolution can reach pixel scale, which provides a useful tool for accurate flow field measurements \cite{SHI_2014}.

In summary,  an ultrasound particle image velocimetry algorithm is proposed to effectively measure the two-dimensional flow field distribution of highly concentrated sand-bearing water streams.  The 2D flow field measurement method combined with UPIV measurement system, which can significantly improve the environmental tolerance and accuracy of flow field measurement,  and reduce the computational consumption by pyramidal hierarchical algorithm.  Experiments have been conducted to verify the performance of the algorithm proposed in the paper.  The results demonstrate that UPIV is a promising technique to generate instantaneous two-dimensional fields of flow velocities in sediment-laden fluids \cite{kim04_devel_valid_echo_piv,Shindler_2012}.  The quantitative performance assessment of the algorithm has been made using ultrasound images acquired in seeded flow.  The algorithm accurately measured the 2D flow field in turbid water with 10\textperthousand volume concentration and outperformed the cross correlation method.

\section{Ultrasound image acquisition}
\label{sec:org902e25d}
\begin{figure}[htbp]
\centering
\includegraphics[width=0.8\columnwidth]{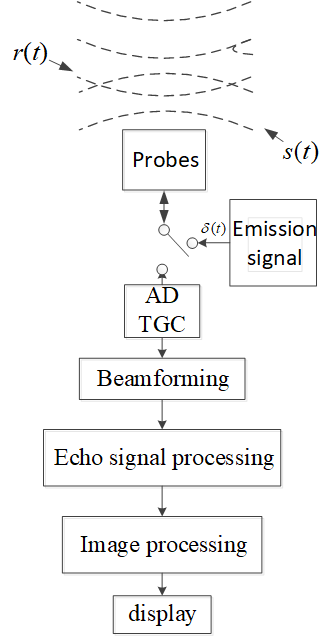}
\caption{\label{fig:org96c253d}
The principle of digital B-mode ultrasound imaging system}
\end{figure}

B-mode ultrasound imaging system, which adopts the ultrasound pulse-echo technique, is widely used.  It transmits ultrasound waves into the medium with an ultrasound transducer.  The waves are reflected, refracted and scattered when meeting objects, and then echo signals with location information are received and proposed by a receiving unit.  Finally, a grayscale image of representing the distribution of objects inside the medium.  The principle is shown in Fig.\ref{fig:org96c253d}.

\begin{figure}[htbp]
\centering
\includegraphics[width=0.3\paperwidth]{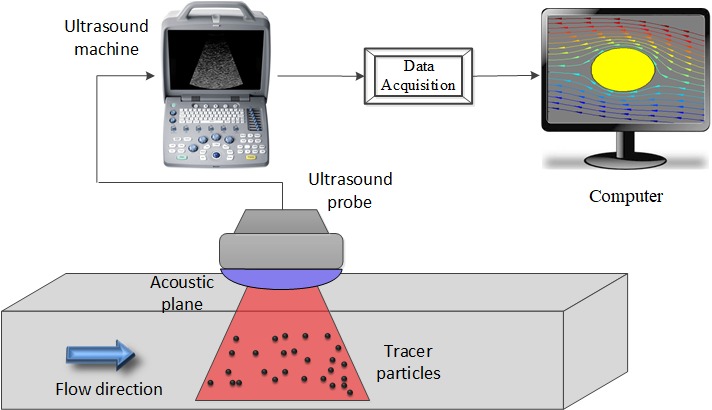}
\caption{\label{fig:org9a898ba}
UPIV system}
\end{figure}

As shown in Fig.\ref{fig:org9a898ba}, an UPIV measurement system is proposed in the paper.  For this system, an acoustic array are motivated to emit acoustic beams.  Those beams point to a specific direction at every launch time \cite{souquet11_ultraf_ultras_imagin}.  After traveling in the water, echoes reflected from tracer particles or suspended sediment are received by the acoustic array \cite{nikolovil_recur}.  The intensity and flight duration of the acoustic echoes are recorded in the system.  Echo intensity reflects the existence and material of the particles in the fluid, and flight duration reflects target distances from the array.  Based on that information, the fluid morphology on a measuring line on depth-direction can be achieved.  Repeat above operations, and a two-dimensional fan-shaped ultrasound image is generated \cite{ng06_model_ultras_imagin_as_linear}.

\section{B-mode ultrasound velocimetry based on optical flow}
\label{sec:orga40bb16}

\subsection{Constraint equation for optical flow algorithm}
\label{sec:orgaeabe8e}

Multiple statinary two-dimensional underwater particle images are captured in a period by the measurement system.  We need to analyze the motion state, \((u, v)\), of the underwater particles in the observed profile at a given moment, where \(u(x, y)\) and \(v(x, y)\) are the velocity components in two dimensions at pixel point \((x, y)\) in underwater images.

In the case of a light source with a constant intensity, the instantaneous change in the grayscale of an image pixel is caused by the mothions of particles.  The image sequence is expressed by \(I(x, y, t)\), \((x, y)\) is a pixel, \(t\) is a time point.  As mentioned above, the instantaneous rate of the change in two dimension at a specific point in underwater profile is expressed by \((u, v)\), which is refered to as the optical flow vector.  Every optical flow vector corresponds to the instantaneous velocity vector at a point, and velocity vectors of all pixels in 2D underwater image constitute the optical flow field. 

Our task is to determine the optical flow field by the variation and correlation of the gray intensity of the pixels in the image sequence in the time domain, so as to obtain the motion of each pixel position, then obtain the 2D optical flow field.

The optical flow calculation usually uses the brightness constancy model (BCM), of which there are two assumption: The brightness is constant between adjacent frames; The objects' motion between adjacent frames is relatively small.

Assuming that \(I(x, y, t)\) is the gray level of pixel \((x, y)\) at the moment \(t\), so the BCM can be expressed by the following equation:

\begin{small}
  \begin{equation}
  \label{eqn:flowbasic}
  I(x+dx, y+dy, t+dt)=I(x, y, t)
  \end{equation}
\end{small}
where, \(dx, dy, dt\) are the increments of \(x, y, t\) respectively.

Expanding \(I(x+dx, y+dy, t+dt)\) with the first-order Taylor formula,  we can obtain following expansion formula:

\begin{strip}
\begin{equation}
\label{eqn:talorExp}
I(x+dx, y+dy, t+dt)=I(x, y, z)+\frac{\partial I}{\partial x}dx+\frac{\partial I}{\partial y}+\frac{\partial I}{\partial t} +e
\end{equation}
\end{strip}

Subtituting Eq. (\ref{eqn:talorExp}) into Eq. (\ref{eqn:flowbasic}), the constraint equation for optical flow algorithm is obtained as below:

\begin{equation}
\label{eqn:flowEqn}
I_xu+I_yv+I_t=0
\end{equation}
where \(u(x, y), v(x, y)\) is the components of optical flow at pixel \((x, y)\), \(I_x, I_y, I_t\) are the partial derivatives of \(I\) with respect to \(x, y, t\), respectively.

\subsection{L-K smoothness constraint equation}
\label{sec:orge6f0fc4}
Equation. (\ref{eqn:flowEqn}) has two variables to be solved, so the solution is not unique.  Constraints should be imposed to get the appropriate optical flow vector.

A predominant constraint method is the L-K algorithm with the consistency constraint assumption.  The constraint of the L-K algorithm is that the pixel motion is consistent in the local region \cite{Fleet_1990}.  Assuming that the motion vector remains constant in the spatial neighborhood of the pixel, weighted least-squares is used to calculate optical flow by assigning different weights to the residual values of different pixel points.  So equation (\ref{eqn:flowEqn}) is rewritten to the following formula:

\begin{tiny}
\begin{equation}
\label{eqn:weightLS}
e(x, y) = \sum\limits_{(x, y)\in \Omega}W^2(x, y)(I_xu+I_yv+I_t)^2
\end{equation}
\end{tiny}
where \(W(x, y)\) is the weight assigned to the residual value of a pixel point, \((x, y)\).

In order to find the optimal value of optical flow to make \(e(x, y)\) minimal, let the partial derivative of \(e(x, y)\) with respect to \(V\)(\(V=(u, v)\)), be 0.  According to the method, the optical flow vector, \(V=(u, v)^T\), can be derived:

\begin{equation}
\label{eqn:LKFlow}
  V=(A^TW^2A)^{-1}A^TW^2b
\end{equation}
where, \(A\) is the gradient matrix of the image \(I(x, y)\):

\begin{tiny}
\begin{equation}
\label{eqn:hessianI}
A = \left[
\begin{array}{cccc}
\nabla I(x_1, y_1) & \nabla I(x_1, y_2) & \ldots & \nabla I(x_1, y_N) \\
\nabla I(x_2, y_1) & \nabla I(x_2, y_2) & \ldots & \nabla I(x_2, y_N) \\
\vdots & \vdots & \ddots & \vdots\\
\nabla I(x_N, y_1) & \nabla I(x_N, y_2) & \ldots & \nabla I(x_N, y_N) \\
\end{array}
\right]
\end{equation}
\end{tiny}
\(W\) is the weight matrix:

\begin{tiny}
\begin{equation}
\label{eqn:weightI}
W=\left[
\begin{array}{cccc}
w(x_1, y_1) & w(x_1, y_2) & \ldots & w(x_1, y_N) \\
w(x_2, y_1) & w(x_2, y_2) & \ldots & w(x_2, y_N) \\
\vdots & \vdots & \ddots & \vdots \\
w(x_N, y_1) & w(x_N, y_2) & \ldots & w(x_N, y_N)
\end{array}
\right]
\end{equation}
\end{tiny}
\(b\) is the partia derivative of \(I(x, y)\) with respect to t:

\begin{tiny}
\begin{equation}
\label{eqn:partialTI}
b=-\left[
\begin{array}{cccc}
I_t(x_1, y_1) & I_t(x_1, y_2) & \ldots & I_t(x_1, y_N) \\
I_t(x_2, y_1) & I_t(x_2, y_2) & \ldots & I_t(x_2, y_N) \\
\vdots & \vdots & \ddots & \vdots \\
I_t(x_N, y_1) & i_t(x_N, y_2) & \ldots & I_t(x_N, y_N)
\end{array}
\right]
\end{equation}
\end{tiny}

According to Eq. (\ref{eqn:LKFlow}), we can calculate the optical flow vectors of the pixels in the acoustic images, from which two dimensional flow velocity field will be inferred.

\subsection{Pyramidal L-K algorithm}
\label{sec:orge61e815}

The constraints of the L-K algorithm are small motion speed, constraint brightness, and consistent region, which are stronger assumptions and not easily satisfied in many applications.  When particles in water are moving fast, the assumptions above do not hold, and computational errors will arise, which would gradually accumulate, making subsequent calculation more deviant and eventually failing to obtain valid and reliable optical flow values.

Because fast motion speed will generate large errors of the algorithm, it is desirable to reduce the motion speed of particles in the ultrasound images to meet the algorithm requirement \cite{bailer15_flow_field}.  An intuitive way to achieve this is to shrink the size of the images.  Suppose that when an image's size is \(400\times 400\), a pixel's flow velocity is 16 pixels per frame; then when the image is downscaled to \(200\times 200\), the pixel's flow velocity is 8 pixels per frame.  Therefore, if a ultrasound image does not satisfy the motion speed assumption, L-K algorithm can still be applied after downscaling the image several times.

\begin{figure}[htbp]

\includegraphics[width=.9\linewidth]{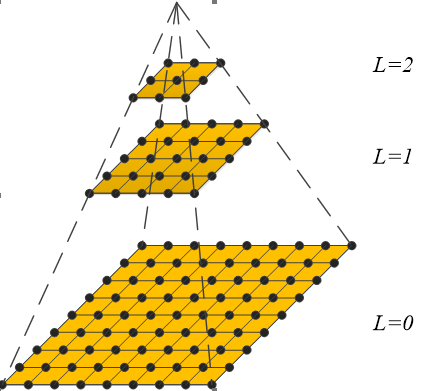}
\caption{\label{fig:org68656d2}
Pyramid images schematic}
\end{figure}

In this way,  a multi-scale representation of images, image pyramid, can be used in the L-K optical flow algorithm, which generates pyramid images of the ultrasound images with different sizes, to solve the problem mensioned above layer by layer.   The basic approach is to obtain the UPIV image sequence, \(I(x, y, t)\), and downsample it separately until it reaches the specified level.  In practice, a pyramid of three or four layers is usually sufficient to meet ther requirements of efficiency and accuracy.  Figure \ref{fig:org68656d2} shows the pyramid images.

The Pyramid L-K algorithm is an iterative calculation in layers, which continuously improves the resolution of the results. The optical flow value calculated from the L layer image is used as the initial value of the optical flow of the L-1 layer.  Then the exact value of the optical flow of the L-1 layer is calculated, which is used as the initial value of the optical flow value of the L-2 layer.  This process continues iteratively until the exact value of the optical flow of the original image is calculated.

\subsection{Implementation of the proposed algorithm}
\label{sec:org19ec479}

The proposed algorithm of 2D flow field measurement of sediment-laden flow based on Pyramid L-K algorithm has three main steps: building pyramid images, tracking pixels through pyramid images, and optical flow calculation iteration.

\subsubsection{Building pyramid images}
\label{sec:orga986cad}

Let \(I_0\) be the image of layer 0 of the original image \(I\), which is the higherst resolution image in the pyramid image, and the width and height of the image are defined as \(W, H\).  To build the pyramid images by downsample by \(1/2\).  Let \(L=0, 1, 2, \cdots\) denote the number of layers of the pyramid: \(I^L\) denotes the image of the Lth layer, then the image of the Lth layer, then the image of the Lth layer of the pyramid can be obtained by the following equation:

\begin{strip}
\begin{equation}
\label{eqn:PyraImgs}
\begin{array}{llll}
I^L(x, y) & = & \frac{1}{4} & I^{L-1}(2x, 2y)+ \\
\specialrule{0em}{1pt}{1pt}
& & \frac{1}{8} [ & I^{L-1}(2 x-1,2 y)+I^{L-1}(2 x+1,2 y)+ \\
& & & I^{L-1}(2 x, 2 y-1)+I^{L-1}(2 x, 2 y+1)] + \\
\specialrule{0em}{1pt}{1pt}
& & \frac{1}{8} [ & I^{L-1}(2 x-1,2 y-1)+I^{L-1}(2 x+1,2 y+1)+\\
& & & I^{L-1}(2 x+1,2 y-1)+I^{L-1}(2 x-1,2 y+1) ]
\end{array}
\end{equation}
\end{strip}

In practice, it is usually sufficient to build 4 layers of pyramid images to meet the application requirements.

\subsubsection{Tracking pixels through pyramid images}
\label{sec:org885391d}

The purpose of image point tracking is to find the matcing pixel point \(N\) in the image to be matched for the pixel point \(M\) in the specified reference image \(I\), or to calculate the optical flow \(V\) fo the point \(M\), where \(V=(u, v)^T\).  Assuming that the coordinates of the point \(M\) in the original image \(I\) are \((M_x, M_y)\), and \(M_L=(M^L_x, M^L_y)\) is the value of the coordinates of the point \(M\) in the \(Lth\) level pyramid image, \(M^L=M/2^L\), which is known from the principle of image pyramid formation \cite{kim07_real_time}.

Let the optical flow prediction in layer \(L\) be \(g^L=[g_x^L, g_v^L]^T\), and remaining optical flow value, optical flow residual displacement vector, be \(d^L=[d_x^L, d_v^L]^T\).  The relationship between the optical flow vector in layer \(L\) and the optical flow vector prediction in layer \(L-1\) can be expressed as follows:

\begin{equation}
\label{eqn:resiFlow}
g^{L-1}=2(g^L+d^L)
\end{equation}

Considering the top layer, layer \(L_m\), has no reliable optical flow prediction, set \(g^{L_m}=[0, 0]^T\), so the optical flow of original image, layer 0, can inferred as follows:
\begin{equation}
\label{eqn:botFlow}
\begin{array}{ll}
d &=g^0+d^0 \\
  &=2(g^1+d^1)+d^0 \\
  &=\cdots \\
  &=\sum\limits_{L=0}^{L_m}2^Ld^L
\end{array}
\end{equation}

As can be seen from Eq. (\ref{eqn:botFlow}), there is an obvious advantage on the pixel tracking of pyramid images, that the residual displacement vector at every level, \(d^L\), is always a very small value when calculating the optical flow vector.

\begin{strip}
\subsubsection{optical flow calculation iteration}
\label{sec:org7244127}

The relationship between the matching error \(\epsilon^L\) and the residual value of optical flow \(d^L=[d_x^L, d_y^L]^T\) in the \(L\) layer can be expressed as follows:

\begin{equation}
\label{eqn:errfunc}
\epsilon^{L}(d_x^L, d_y^L)=\sum\limits_{x, y}(I(x, y)-J(x+g_x^L+d_x^L, y+g_y^L+d_y^L))^2
\end{equation}
\begin{equation}
\label{eqn:resflowL}
d^L = [d_x^L, d_y^L]^T = arg\, min\{\epsilon^L\}
\end{equation}

let \(A(x, y)=i^l(x,y)\), \(B(x, y)=j^l(x, y)\), then equation (\ref{eqn:errfunc}) can be expressed as:

\begin{equation}
\label{eqn:errfuncab}
\epsilon(d_x^l, d_y^l)=\sum\limits_{x, y}(A(x, y)-B(x+d_x^l, y+d_y^l))^2
\end{equation}

Solving the above equation using the L-K algorithm, the first-order partial derivative of \(\epsilon\) with respect to \((d_x^L, d_v^L)\) is zero when \(\epsilon\) is the minimum, and we have:

\begin{equation}
\label{eqn:epsPartD}
\left.\frac{\partial\epsilon}{\partial d^L}\right\rvert_{d^L=d^L_{opt}}=[0, 0]
\end{equation}

Subtituting Eq. (\ref{eqn:errfuncab}) into Eq. (\ref{eqn:epsPartD}) yields:

\begin{equation}
\label{eqn:epsPartDExp}
\frac{\partial \epsilon}{\partial d^L}=-2\sum\limits_{x, y}(A(x, y)-B(x+d^L_x, y+d^L_y))\left[\frac{\partial B}{\partial x}\quad \frac{\partial B}{\partial y}\right]
\end{equation}

Expanding Eq. (\ref{eqn:epsPartDExp}) with the first-order Taylor formula yields:

\begin{equation}
\label{eqn:BTalorExp}
B(x+d^L_x, y+d^L_y)=B(x, y)+\left[\frac{\partial B}{\partial x} \quad \frac{\partial B}{\partial y}\right]\cdot\left[d^L_x, d^L_y\right]^T+o(x, y)
\end{equation}

Subtituting Eq. (\ref{eqn:BTalorExp}) into Eq. (\ref{eqn:epsPartDExp}) yields:

\begin{equation}
\label{eqn:epsAppr}
\frac{\partial \epsilon}{\partial d^L}\approx -2 \sum\limits_{x, y}(A(x, y)-B(x, y)-\left[\frac{\partial B}{\partial x}\quad \frac{\partial B}{\partial y}\right]\cdot d^L)\left[\frac{\partial B}{\partial x}\quad \frac{\partial B}{\partial y}\right]
\end{equation}

The value of \(A(x, y)-B(x, y)\) can be considered as the derivative of the ultrasound image at the point \((0, 0)\) with respect to time, denoted as:

\begin{equation}
\label{eqn:pixelPartial}
\delta I(x, y)=A(x, y)-B(x, y)
\end{equation}

The gradient vector of the ultrasound image can be defined as:

\begin{equation}
\label{eqn:gradImg}
\nabla I=\left[\begin{array}{c}
I_x\\
I_y
\end{array}\right]=\left[
\frac{\partial B}{\partial x}\quad \frac{\partial B}{\partial y}\right]^T
\end{equation}

The gradient vector \([I_x, I_y]\) of the image \(A(x, y)\) can be calculated from the neighborhood of the image at the pixel \(M\).  The gradient of teh image can be expressed in terms of the central difference as:
\begin{equation}
\label{eqn:gradImgXY}
\begin{split}
I_x(x, y)=\frac{\partial A(x, y)}{\partial x}=\frac{A(x+1, y)-A(x-1, y)}{2}\\
I_y(x, y)=\frac{\partial A(x, y)}{\partial y}=\frac{A(x, y+1)-A(x, y-1)}{2}
\end{split}
\end{equation}

Subtituting Eq. (\ref{eqn:gradImg}) and Eq. (\ref{eqn:gradImgXY}) into Eq. (\ref{eqn:epsAppr}), we get:
\begin{equation}
\label{eqn:epsPartDI}
\begin{array}{lll}
\frac{1}{2}\frac{\partial \epsilon}{\partial d^L} & \approx & (\nabla I^T\cdot d^L-\delta I)\nabla I^T \\

& = & \sum\limits_{x, y}\big(\left[
  \begin{array}{cc}
    I_x^2 & I_xI_y \\
    I_xI_y & I_y^2
  \end{array}\right]-\left[
  \begin{array}{c}
    \delta I\cdot I_x\\
    \delta I\cdot I_y
  \end{array}\right]\big) \\

& = & G\cdot d^L-b
\end{array}
\end{equation}
where, 

\begin{equation}
\label{eqn:GbForm}
  G = \sum\limits_{x, y}\left[
      \begin{array}{cc}
        I_x^2 & I_xI_y \\
        I_xI_y & I_y^2
      \end{array}\right], \quad
  b = \left[
      \begin{array}{c}
        \delta I\cdot I_x\\
        \delta I\cdot I_y
      \end{array}\right]
\end{equation}

When the error \(\epsilon\) achieves its minimum value, the optimal optical flow residual displacement vector \(d_{opt}^L\) of the \(L\) layer can be derived as follows:

\begin{equation}
\label{eqn:optOptFlow}
d_{opt}^L=G^{-1}b
\end{equation}

To further solve the optical flow residual displacement vector \(d^L\) in the \(L\) layer accurately, the Newton-Raphson method can be used to improve the exact solution of the optical flow.

Let the optical flow value after \(k-1\) iterations at the \(L\) layer be \(\zeta^{k-1}=[\zeta_x^{k-1}, \zeta_y^{k-1}]^T\).  Then, when the \(k\)th iteration is performed, the image after superimposing a translation of the vector \(\zeta^{k-1}\) is:

\begin{equation}
\label{eqn:ImageTrans}
B_k(x, y)=B(x+\zeta_x^{k-1}, y+\zeta_y^{k-1})
\end{equation}

The objective of the \(k\)th iteration is to calculate the optical flow residual displacement vector \(\lambda^k=[\lambda_x^k, \lambda_y^k]^T\) at the \(L\) layer, which minimizes the error function represented as follows:

\begin{equation}
\label{eqn:epsErrFun}
\epsilon(\lambda_x^k, \lambda_y^k)=\sum\limits_{x, y}(A(x, y)-B(x+\lambda_x^k, y+\lambda_y^k))^2
\end{equation}  

Solving the above equation,  we get:

\begin{equation}
\label{eqn:iterOptFlow}
\lambda^k=G^{-1}b_k
\end{equation}
where, \(G\) remains constant in the iteration, and 

\begin{equation}
\label{eqn:deltaB}
\begin{array}{l}
b=\sum\limits_{x, y}\left[\begin{array}{c}
\delta I_k\cdot I_x \\
\delta I_k\cdot I_y
\end{array}\right]
\delta I_k(x, y)=A(x, y)-B_k(x, y)\\
\end{array}
\end{equation}

From the residual displacement vector, \(\lambda^k\), calculated from Eq. (\ref{eqn:epsErrFun}), the displacement, \(\zeta^k\)  of the ultrasound image at the \(k\)th iteration can be deduced as:

\begin{equation}
\label{eqn:iterOper}
\zeta^k=\zeta^{k-1}+\lambda^k
\end{equation}

The process is continued until the calculated residual displacement vector is less than a given threshold, or the maximum number of iteration is reached.

Suppose the convergence condition of Eq. (\ref{eqn:iterOper}) is reached after \(k\) iterations, then the final optical flow of the \(k\) layer can be expressed as:

\begin{equation}
\label{eqn:finalOptFlow}
d^L=\zeta^N=\sum\limits_{k=1}^K \lambda^k
\end{equation}

The corresponding optical flow vectors are obtained by using the above method in all layers of the pyramid images.  Substituting them back into Eq. (\ref{eqn:botFlow}), the total optical flow vectors can be obtain.
\end{strip}

\section{Experimental Setup \label{sec:Exp}}
\label{sec:orgdf2bf4f}
\subsection{Water Flow Experimental conditions}
\label{sec:org29093cf}

\begin{figure}[htbp]
\centering
\includegraphics[width=0.3\paperwidth]{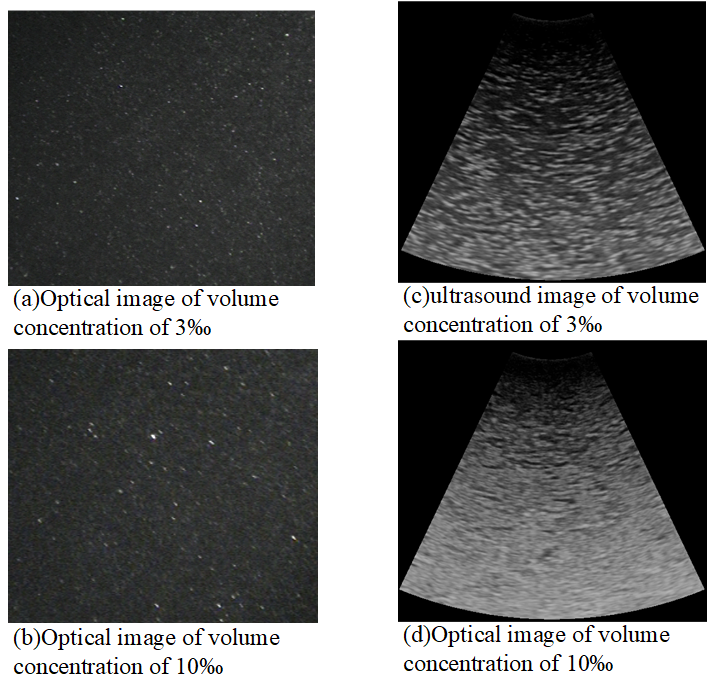}
\caption{\label{fig:orga99d718}
Comparison of optical images and ultrasound images}
\end{figure}

Since the wavelength of ultrasound, \(\lambda_u\) is in the range of \(0.15\sim 1.5mm\), and the wavelength of visible light, \(\lambda_o\) is in the range of \(380\sim 780nm\), \(\lambda_u\gg\lambda_o\), the spot of the particle image in ultrasound imaging is larger than it in optical imaging, as shown in Fig. \ref{fig:orga99d718}.  For this reason, the particle images in ultrasound imaging are more likely to saturate.  In general, as shown in Fig. \ref{fig:orga99d718} (d), ultrasound images will basically reach saturation when the volume concentration of sand content in the water stream reaches \(10\)\textperthousand, namely mass concentration is about \(10\,kg/m^3\).

In order to ensure the smoothness of the experiment, the measurements are conducted in a constant sand-bearing water flow with a volume concentration of \(10\)\textperthousand, the water depth is about \(50\, cm\), and the range of width-to-depth ratio is \(1.2\sim1.5\), which ensures that the side walls of the flume will not produce obvious secondary turbulence. The experimental observation area is about \(2\, m\) away from the inlet of the flume, which ensures that the high concentration of sand-bearing water has been fully spread when arriving the observation area.

\subsection{Water tank experimental environment and facilities}
\label{sec:orgc0a0c37}

As shown in Fig. \ref{fig:org9a898ba}, a variable-slope recirculating water tank is used for the experiments. The sink is \(300\, cm\) long, \(50\, cm\) wide, and \(50\, cm\) deep, and the inner cavity is divided into left and right parts by a vertical glass plate.  A grid is equipped in the middle of the flume to rectify and reduce the fluctuation of the water surface.  An inlet pipe and a drainage pipe are equipped at one end of the tank, which are connected to a pump.  The high concentration sand-containing water flow is driven by the pump, enters the tank from a reservoir through the inlet pipe.  The water flow circulates and then discharges back to the reservoir through the drainage pipe.  The volume concentration of sand-bearing water in experiments is 10\textperthousand, i.e., the mass concentration is about \(10\, kg/m^3\), Reynolds number is 2200, and the average diameter of sand particles is \(100\, \mu m\).

The experimental system controls the speed of the pump with a frequency converter to provide a stable flow environment for the experiment.  The reservoir with a total area of about \(3 m^2\) is built at the end of the tank to regulate the flow pattern and water volume for the high concentration sand-bearing water flow experiments, while a flexible rubber tube is used to connect the tank to the pump motor to reduce the impact of the pump motor vibration on the measurement.

\subsection{Ultrasonic Water Flow Imaging System}
\label{sec:org219f9a7}

In the experiments, a medical B-type ultrasound instrument, SIUI APOGEE 1200, is deployed. The ultrasound probe is fixed just touching the water surface to avoid affect the flow field.  Emission frequency of the probe is 5.0 MHz, and the monitoring range of the probe is \(10\times35\, cm^2\).  The imaging size of the ultrasound instrument is \(640\times 480\) pixels, the spatial resolution is about \(32\times32\, pixels/cm^2\), and the temporal resolution is \(60\, frames/second\). In order to monitor the motion of all particles in the imaging plane, the probe is placed in the same direction as the sand-containing water flow, and the ultrasound instrument is connected to a computer, which converts the echo signals collected by the probe into ultrasound images and transmits them to the ultrasound particle velocimetry software in the computer for analysis, processing, and display of flow field information.

The experimental steps are as below.  Firstly, adjust the inclination of the sloping water tank and the speed of the pump motor, so that the water depth meets the measurement requirements.  Then adjust the direction of the ultrasonic probe with a laser collimator to be the same as the flow direction of the sand-containing water.  Add appropriate amount of sand particles as tracer particles, and wait about 10 minutes so the various conditions of the whole measurement system are stable.  At this time, ultrasonic imaging of the water flow is performed to obtain ultrasound particle images.  Finally, the ultrasonic particle images were analyzed using the algorithm proposed in this paper to obtain two-dimensional flow field of turbid water stream.

\section{Result and error analysis \label{sec:Con}}
\label{sec:orgdf919f0}

The algorithm proposed in the paper is applied to the acquired ultrasonic underwater images to calculate the underwater two-dimensional flow field to verify the effectiveness of the algorithm.

\begin{figure}[htbp]
\centering
\includegraphics[width=0.3\paperwidth]{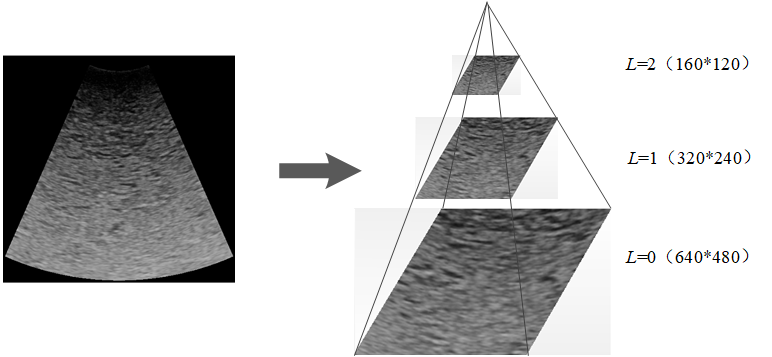}
\caption{\label{fig:org0018b0d}
Pyramidal images of high concentration ultrasound images (3 layers)}
\end{figure}

The original ultrasound particle image of \(640\times480\) pixels is downsampled by 2:1 to obtain the image of the next layer.  Bilinear interpolation method is used between adjacent layers to make the transition smoother.  The obtained image sequence of the pyramid structure is shown in Fig. \ref{fig:org0018b0d}.

\begin{strip}
The experimental results are measured by Average Angle Error (AAE) and Standard Angle Error (SAD).  The average angle error is calculated as:

\begin{equation}
\label{eqn:AAE}
\mathbf{AAE}= \frac{1}{N}\sum\limits_{i=1}^N\phi_e(i)
\end{equation}
where:
\begin{equation}
\label{eqn:phiE}
\phi_{e}(i)=\arccos \left[\frac{u_{i}^{c} u_{i}^{e}+v_{i}^{c} v_{i}^{e}+k^{2}}{\left(u_{i}^{c}\right)^{2}+\left(v_{i}^{c}\right)^{2}+k^{2} \sqrt{\left(u_{i}^{e}\right)^{2}+\left(v_{i}^{e}\right)^{2}+k^{2}}}\right]
\end{equation}
\end{strip}
where N denotes the total number of pixels in a frame of ultrasound particle image, \((u_i^c, v_i^c)\) denotes the ground truth value of the velocity of the \(i\)th pixel, i.e., standard optical flow value, \((u_i^e, v_i^e)\) denotes the optical flow vector value of the \(i\)th pixel, and \(k\) denotes the number of frame intervals between images.  \(\phi_i(i)\) denotes the angular error of the \(i\)th pixel, and \(\mathbf{AAE}\) denotes the average angular error of the entire optical flow field.

The average angular error reflects the deviation of the optical flow vector field calculated using the optical flow algorithm from the standard vector field. In the experiments, the frame interval \(k\) is set as 1, and the velocity measured by a high-precision acoustic doppler velocimetry (ADV) is used as the standard optical flow value \(u_i^e, v_i^e\).

The standard angle error can be expressed as:

\begin{equation}
\label{eqn:sad}
\mathbf{SAD} = \sqrt{\frac{1}{n}\sum\limits_{i=1}^n(\phi_e(i)-\mathbf{AAE})^2}
\end{equation}

The standard angular error reflects the fluctuation of \(\mathbf{AAE}\) in the calculated optical flow vector field.

\begin{figure}[htbp]
\centering
\includegraphics[width=0.3\paperwidth]{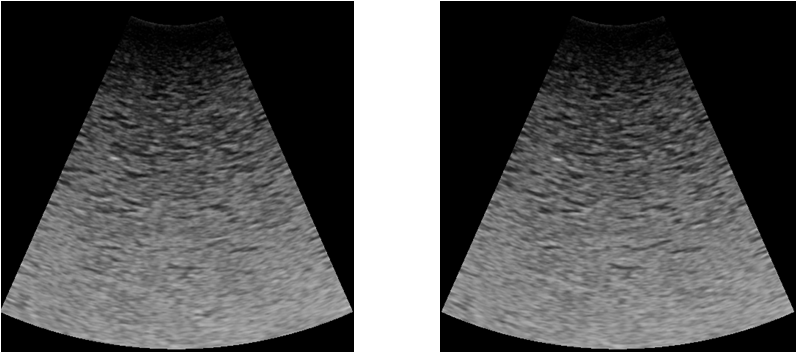}
\caption{\label{fig:orgbb04c2e}
The sequence of ultrasound particle images with a volume concentration of 10\textperthousand and a flow velocity of $35\, cm/s$}
\end{figure}

\begin{figure}[htbp]
\centering
\includegraphics[width=0.3\paperwidth]{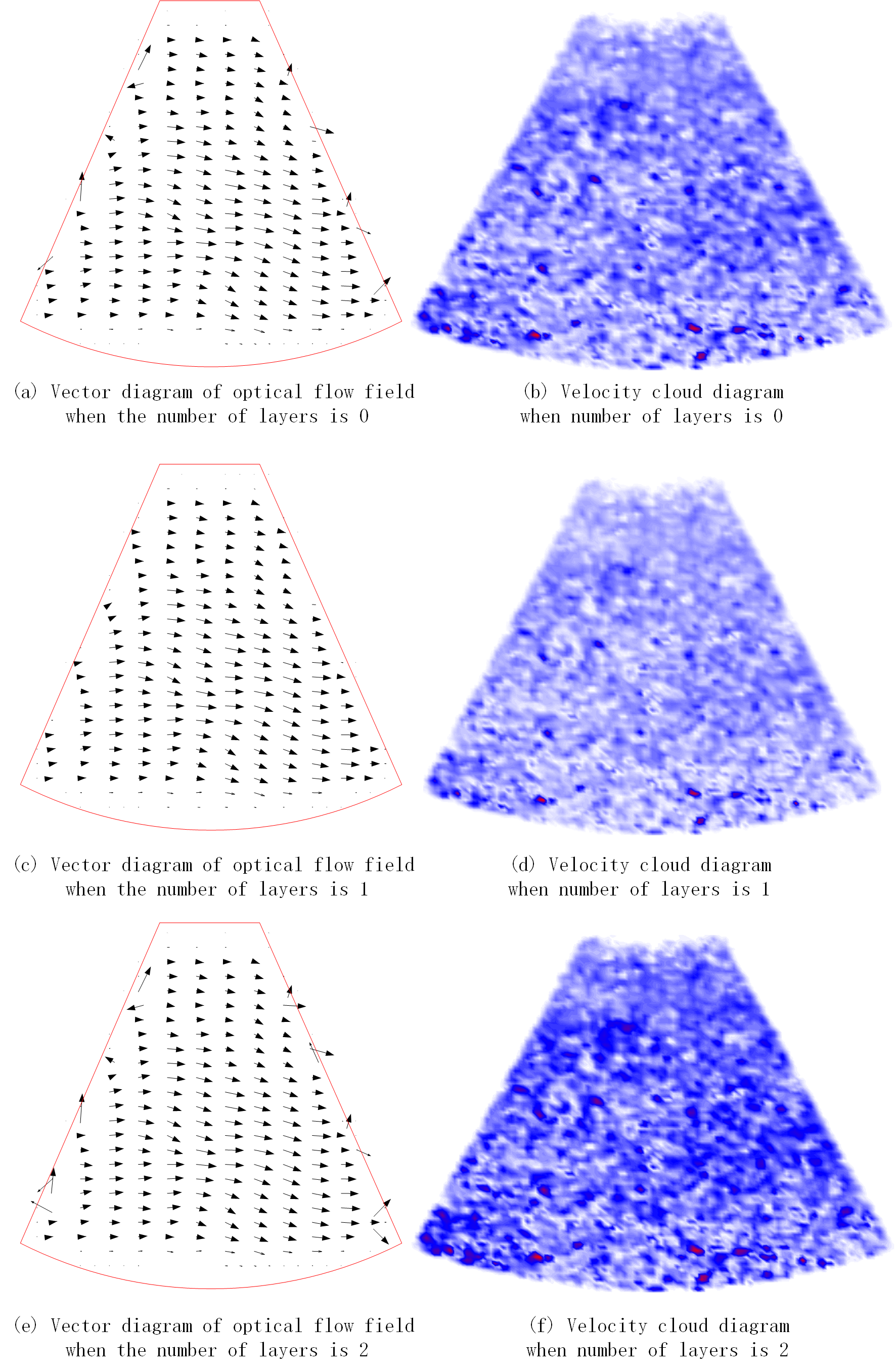}
\caption{\label{fig:org7c396ec}
The optical flow field and velocity clouds with a volume concentration of 10\textperthousand and a flow velocity of $35\, cm/s$}
\end{figure}

For verifying the effectiveness of the Pyramid L-K optical flow algorithm in the case of large flow velocities, experiments are conducted using a sequence of ultrasonic particle images with a volume concentration of 10\textperthousand and a flow velocity of \(35\, cm/s\) for experimental analysis, as shown in Fig. \ref{fig:orgbb04c2e}. Figure \ref{fig:org7c396ec} (a1)\(\sim\)(c1) shows the optical flow field obtained using the Pyramid L-K optical flow algorithm for the number of stratifications 0, 1, and 2, and Figure \ref{fig:org7c396ec} (a2)\(\sim\)(c2) shows the velocity clouds corresponding to the optical flow field.

From Fig. \ref{fig:org7c396ec} (a1)\(\sim\)(c1), we can see that the Pyramid L-K optical flow algorithm can calculate the optical flow at each point in the image observation area.  However, with the increase of the number of layers, the optical flow calculation at the boundary of the observation area becomes worse , which is because the L-K optical flow algorithm uses local constraints so that the calculation of the optical flow at each pixel needs to depend on the optical flow values of other pixels in the neighborhood.  While the grayscale values at the edge of the image are zero, the optical flow in those regions can not be calculated accurately.

\begin{table}
  \centering
  \caption{Comparison of optical flow calculation error of ultrasound particle image sequence with volume concentration of 10\textperthousand and flow rate of   latex $35\, cm/s$ }
  \label{tab:CompRslt}
   \begin{tabular}{cccc}
      \toprule
      Layer & AAE & SAD & Time-consuming \\
      \midrule
      0 & 4.87 & 7.13 & 56 \\
      1 & 3.74 & 5.64 & 29 \\
      2 & 5.43 & 8.67 & 16 \\
      \bottomrule
  \end{tabular}
\end{table}

Table \ref{tab:CompRslt} shows the mean angular error, standard angular error and operation time of the optical flow vector field calculated by the proposed ultrasound image velocimetry algorithm.  From the results, it can be seen that the mean angular error decreases from 4.87 to 3.74 and the standard angular error decreases from 7.13 to 5.64 when the number of stratification increases from 0 to 1.  When the number of stratification increases from 1 to 2, the mean angular error increases from 3.74 to 5.43 and the standard angular error increases from 5.64 to 8.67, which is because the basic constraints of the L-K algorithm are: motion velocity is  small and the local pixel motion is consistency.

\begin{figure}[htbp]
\centering
\includegraphics[width=0.3\paperwidth]{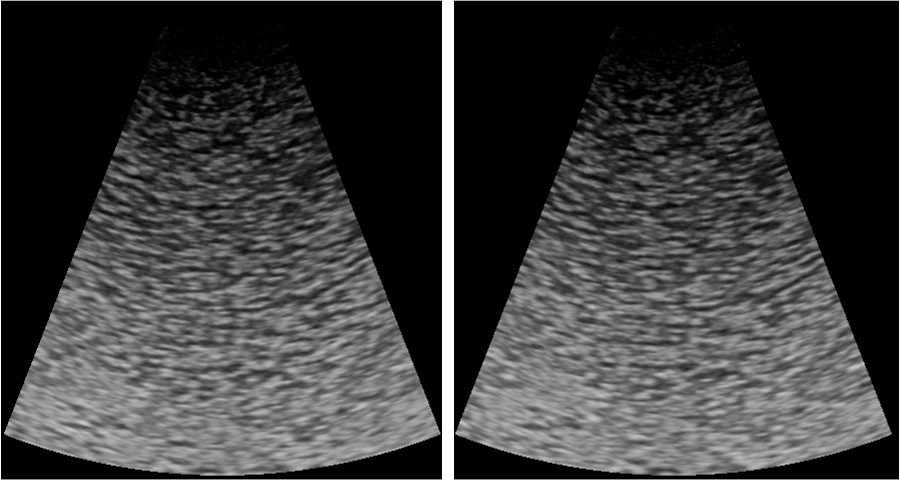}
\caption{\label{fig:org400f13c}
Ultrasonic particle image sequence with volume concentration of 10\textperthousand and flow rate of $25\, cm/s$}
\end{figure}

\begin{figure}[htbp]
\centering
\includegraphics[width=0.4\paperwidth]{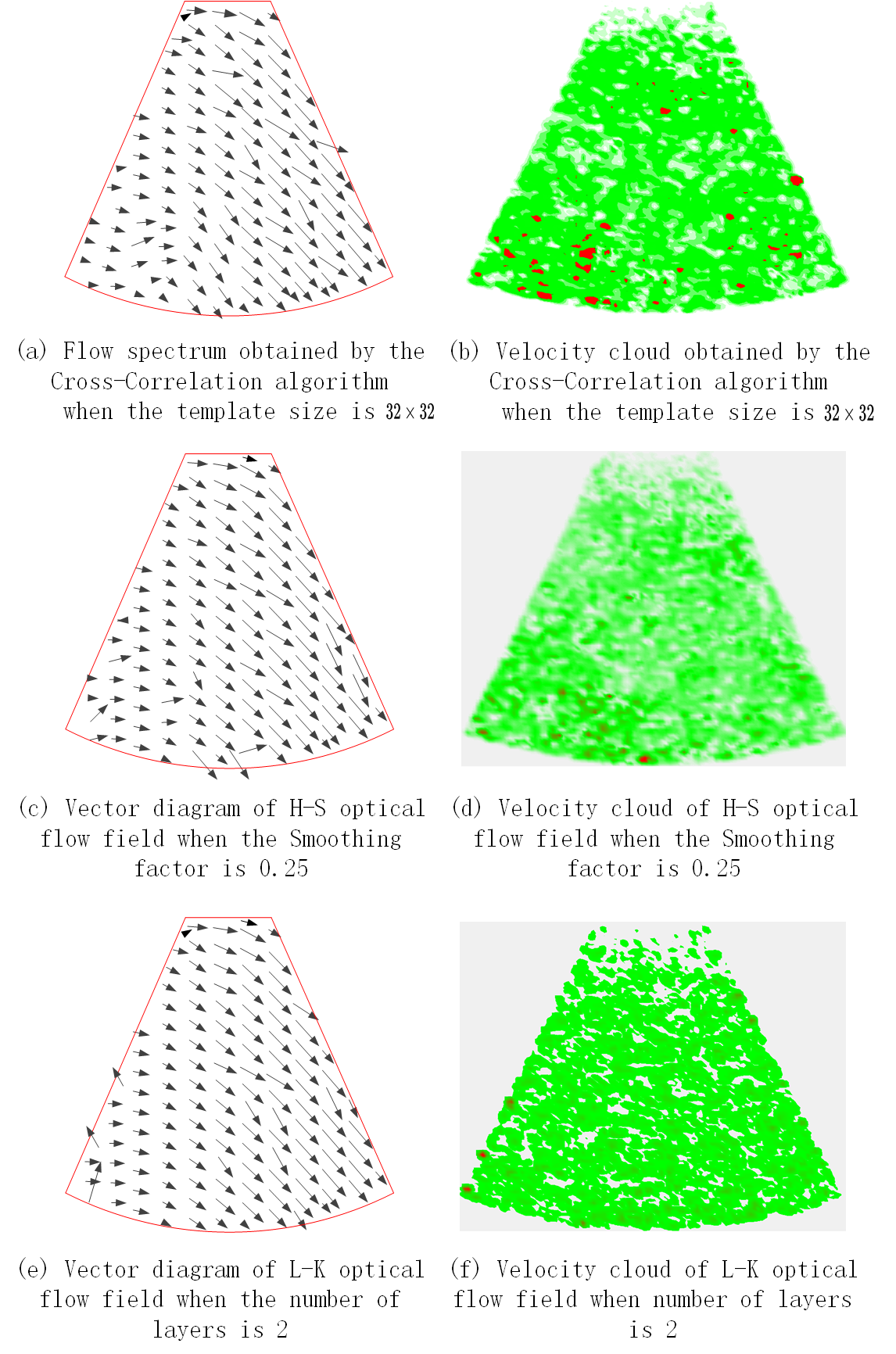}
\caption{\label{fig:org7eda855}
Flow spectra and velocity clouds derived by applying Pyramid L-K optical flow and Cross-Correlation algorithm to ultrasound particle image sequence with volume concentration of 10\textperthousand and flow velocity of $25\, cm/s$}
\end{figure}

The performance of Pyramid L-K optical flow and the traditional algorithm, Cross-Correlation algorithm, are compared horizontally to further validate the performance of the proposed algorithm.  The experiments choose a high concentration ultrasonic particle image sequence with a volume concentration of 10\textperthousand and a flow rate of \(25\, cm/s\) for experimental analysis, as shown in Fig. \ref{fig:org400f13c}.  The experiments are conducted with typical templates with sizes of \(32\times32\), \(48\times48\), and \(64\times64\) pixels are selected for matching in the template matching-based mutual correlation algorithm.  The number of stratifications of 0, 1, and 2 are selected for comparative analysis in the Pyramid L-K optical flow-based algorithm, as shown in Tab. \ref{tab:perfComp}.  Figure \ref{fig:org7eda855} (a1)\(\sim\)(b1) shows the velocity field obtained using the Pyramid L-K optical flow, Cross-Correlation algorithm, and Figure \ref{fig:org7eda855} (a2)\textasciitilde{}(b2) shows the velocity cloud maps corresponding to the velocity fields.

\begin{table}[!t]
  \caption{Performance comparison of Pyramid L-K, Cross-Correlation algorithms}
  \label{tab:perfComp}
  \begin{tabular}{llccc}
    \toprule
    algo- & para- & AAE & SAD & time-  \\
    rithm & meter &     &     & consuming  \\
    \midrule
    cross-            & 32$\times$32 & 12.14 & 21.04 & 6.2 \\
    correlation       & 48$\times$48 & 14.85 & 23.86 & 5.5 \\
                      & 64$\times$64 & 15.73 & 24.43 & 4.7 \\
    H-S               & \(\alpha=0.25\) & 3.92 & 9.11 & 66 \\
    algorithm         & \(\alpha=0.5\)  & 5.31 & 8.54 & 62 \\
                      & \(\alpha=0.75\)  & 7.04 & 7.99 & 65 \\
    Pyramid           & Layer 0    & 4.59  & 6.85  & 54  \\
    L-K               & Layer 1    & 2.43  & 5.33  & 28  \\
                      & Layer 2    & 5.12  & 8.39  & 17  \\
    \bottomrule
  \end{tabular}
\end{table}

The velocity vector fields obtained after processing by each algorithm are shown in Fig. \ref{fig:org7eda855} (a1)\(\sim\)(c1), from which it can be intuitively seen that the flow spectra obtained by the H-S optical flow and Pyramid L-K optical flow based algorithms are better and the flow fields are smoother with only a few error matches, while the results obtained by the cross-correlation algorithm based on template matching have more error vectors and the velocity vector fields are not smooth enough.  On the other hand, Figure. \ref{fig:org7eda855} (a2)\textasciitilde{}(c2) qualitatively compare the errors of H-S optical flow, Pyramid L-K optical flow, and Cross-Correlation algorithm.  From that it can be seen that the velocity cloud map of the cross-correlation algorithm has the largest number and area of read spot regions, which reflects the largest error of the algorithm.  It is due to the high concentration of particles in the sand-bearing water flow, which makes the correlation plane emerge with multiple peaks and leads to a large number of erroneous displacement vectors.  There are some small red spots appear in the lower left corner of the velocity cloud from the H-S optical flow algorithm.  It indicates that the error is slightly larger in this area than the others.  Because the concentration of sand in this region is high, the secondary reflection of ultrasound in these places is more serious, and the grayscale of ultrasound image changes more drastically.  Under the condition, the basic constraint of H-S optical flow algorithm, i.e., the assumption of grayscale conservation, not strictly valid, resulting large errors.  The velocity cloud map from the Pyramid L-K optical flow algorithm is the smoothest and with the smallest error.  However, the errors are larger at the edges of the observation area than other areas.  This is because the Pyramid L-K optical flow algorithm uses a local constraint, which makes the calculation of optical flow at each pixel depend on the optical flow values of other pixels in its neighborhood.  The grayscale value at the edges of the ultrasound particle image is zero, which results in a large error in the calculation of the optical flow value at the boundary.

To quantitatively analyze the processing results of various algorithms, table \ref{tab:perfComp} lists the mean angular error, standard angular error and operation time of the H-S optical flow, Pyramid L-K optical flow and FFT Cross-Correlation algorithms.  From the table, it can be seen that the FFT Cross-Correlation is the fastest, but its average angular error and standard angular error are the largest, too.  The H-S algorithm has the longest operation time, but its accuracy is better than FFT Cross-Correlation algorithm.  By adjusting the smoothing factor, the average angular error can be reduced to 3.92.  The Pyramid L-K algorithm has a better calculation efficiency than H-S algorithm, and the best accuracy among the three algorithms. It shows that Pyramid L-K algorithm can achieve the good balance of computational efficiency and accuracy by choosing the appropriate number of layers.

It can also be seen from Tab. \ref{tab:perfComp} that when the template size increases from \(32\times32\) to \(64\times64\), the calculated average angular error does not change much.  Cross-Correlation algorithm is not very sensitive to its parameter.  However, the H-S optical flow and Pyramid L-K optical flow algorithms are particularly sensitive to their parameters.  Therefore, it is necessary to choose carefully when using parameters to keep the error from increasing.

\section{Conclusions \label{sec:Con}}
\label{sec:org95db22e}

This paper describes an ultrasonic PIV algorithm to measure instantaneous two-dimensional velocity field and depth-average velocity for sediment-laden fluid.  Experiments are conducted in turbid and sandy water flow to measure and analyze the flow velocity field with the algorithm proposed in the paper.  From the experimental results, it can be seen that the Pyramid L-K optical flow algorithm can satisfy the optical flow constraint equation even at large flow velocities because of the downsampling of the original image to form a pyramid structure.  The algorithm also adopts the "Coarse to Fine" matching, which finds the large motions in the higher level pyramid images first, and then gradually refines the result, corrects the accuracy of the large motions.  The measure not only improves the computational accuracy but also reduces the computing time. The experimental results prove that the Pyramid L-K optical flow algorithm has the advantages of high computational accuracy, wide range of application, and efficient operation speed.

\end{document}